\documentclass[letterpaper, 10 pt, conference,onecolumn,draftclsnofoot]{ieeeconf}
\IEEEoverridecommandlockouts                              
\usepackage{cite}

\usepackage{array,pdfsync,enumerate,tikz,enumerate}
\usepackage[dvips]{epsfig}
\usepackage{amssymb,amsmath}
\usepackage{caption}
\usepackage{subcaption}
\usepackage{color}
\definecolor{linkcol}{rgb}{0,0.0,1.0}
\definecolor{citecol}{rgb}{0.0,0.6,0.0}

\IEEEoverridecommandlockouts
\usepackage{cite}
\usepackage{amsmath,amssymb,amsfonts}
\usepackage{graphicx}
\usepackage{textcomp}

\usepackage{hyperref}
\hypersetup{colorlinks,breaklinks,
	linkcolor=linkcol,urlcolor=citecol,
	anchorcolor=linkcol,citecolor=citecol,bookmarks=true}

\usepackage{caption}
\usepackage{subcaption}

\newtheorem{theorem}{Theorem}[section]
\newtheorem{proposition}[theorem]{Proposition}
\newtheorem{remark}[theorem]{Remark}
\newcommand{\vx}[1]{\mathbf{#1}}

\usepackage{graphicx}      
\usepackage{cite}
\begin{document}
	
	\title{Quadratic Optimization-Based Nonlinear Control for Protein Conformation 
	Prediction}
	
	\author{Alireza Mohammadi and Mark W. Spong
		\thanks{A. Mohammadi is with the Department of Electrical \& Computer Engineering, University of 
			Michigan-Dearborn, Dearborn, MI 48128 USA. Mark W. Spong is with the
			University of Texas at Dallas, Richardson, TX 75080 USA. Mark W. 
			Spong is the Excellence in Education Chair at the University of Texas at Dallas.  
			Emails: {\tt\small \{amohmmad@umich.edu$^{\ast}$, mspong@utdallas.edu\}}, $^{\ast}$Corresponding Author: A. Mohammadi}%
	}

\maketitle		
		
		\begin{abstract}                
			  \textcolor{black}{Predicting the final folded structure of protein molecules and simulating their folding pathways is of crucial importance for designing viral drugs and studying diseases such as Alzheimer's at the molecular level.  To this end, this paper investigates the problem of protein conformation prediction under the 
			 constraint of avoiding high-entropy-loss routes during folding. Using the well-established kinetostatic compliance (KCM)-based nonlinear dynamics of a protein molecule,  this paper formulates the protein conformation prediction as a pointwise optimal control synthesis problem cast as a quadratic program (QP). It is shown that the KCM torques in the protein folding literature can be utilized for defining a reference vector field for the QP-based control generation problem. The resulting kinetostatic control torque inputs will be close to the KCM-based reference vector field  and guaranteed to be constrained by a predetermined bound; hence, high-entropy-loss routes during folding are avoided while the energy of the molecule is decreased.}
		\end{abstract}
		


\renewcommand\floatpagefraction{.9}
\renewcommand\dblfloatpagefraction{.9} 
\renewcommand\topfraction{.9}
\renewcommand\dbltopfraction{.9} 
\renewcommand\bottomfraction{.9}
\renewcommand\textfraction{.1}   
\setcounter{totalnumber}{50}
\setcounter{topnumber}{50}
\setcounter{bottomnumber}{50}
\newenvironment{psmallmatrix}
{\left(\begin{smallmatrix}}
	{\end{smallmatrix}\right)}
%
    \section{Introduction}
%

Computer-aided prediction of the folded structure of a protein molecule lies at the heart of protein engineering, drug discovery, and investigating diseases such as  Alzheimer's at the molecular and cellular levels~\cite{chirikjian2005analysis}. The 3D structure of a protein molecule,  known as the \textbf{protein conformation}, is mainly determined by its linear amino acid (AA) sequence~\cite{finkelstein2016protein}. The \textbf{protein folding} problem is concerned with determining the final folded structure of a protein molecule given its linear AA sequence.  \textcolor{black}{Studying the folding pathways is also important for designing viral drugs that cause misfolding in virus proteins~\cite{bergasa2020interdiction}.}

Knowledge-based and  physics-based methods are the two prominent computational approaches for solving the protein folding problem.  Knowledge-based methods, which also include the approach of Google's DeepMind AlphaFold~\cite{alquraishi2020watershed}, rely on using previously determined types of folds to solve the protein folding problem~\cite{moult2014critical}. In physics-based methods, on the other hand, protein folding simulations are carried out using the first principles~\cite{oldziej2005physics}.
Physics-based methods enjoy several advantages over their knowledge-based counterparts such as the ability to model the protein-nucleic acid interactions and to  \emph{simulate protein folding pathways}. However, physics-based methods relying on standard molecular dynamics (MD) suffer from numerical instabilities~\cite{lippert2007common}.  To address the inherent challenges of MD-based protein folding solutions, Kazerounian and collaborators~\cite{kazerounian2005nano,kazerounian2005protofold,tavousi2015protofold,Tavousi2016} introduced the kinetostatic compliance method (KCM) to overcome the computational problems arising from large 
degrees-of-freedom (DOFs) and to provide a faster convergence to the molecular minimum energy state. This framework 
assumes that folding takes place as a quasi-equilibrium process where the protein chain complies under the kinetostatic effect of the
inter-atomic and intramolecular force fields. The KCM framework has been successful in investigating 
phenomenon such as the effect of hydrogen bond formation on the protein mechanical mobility~\cite{shahbazi2010hydrogen} 
and conformational trajectory planning of proteins in 3D workspace~\cite{madden2009residue}.

Along various modeling approaches, numerous optimization techniques have also been adopted for obtaining realistic 
folding pathways such as homotopy-based~\cite{dunlavy2005hope} and optimal control-based~\cite{arkun2010prediction,arkun2012combining,arkun2011protein} algorithms. \textcolor{black}{Using a bead-spring linear time-invariant (LTI) model of the protein molecule}, Arkun and Erman in~\cite{arkun2010prediction} propose using a linear quadratic regulator (LQR) control input for guiding the folding simulations in order to simultaneously evade high-energy regions of the folding landscape and to \emph{avoid} high-entropy-loss routes. An interesting aspect of Arkun and Erman's optimal control framework, which is inspired from the behavior of the two-state protein molecules~\cite{weikl2003folding}, is to provide \emph{the optimal solution to protein folding as a trade-off between the molecule energy minimization and the avoidance of high-entropy-loss
routes}. In particular, to avoid extremely fast decay of excess entropy, Arkun and Erman penalize the control input by using an infinite horizon LQR cost function.  Arkun and G{\"u}r in~\cite{arkun2012combining} extended the aforementioned work by providing the sampled trajectories of the coarse-grained protein model as initial conditions for an all-atom MD-based simulation environment, namely, \textcolor{black}{Nanoscale Molecular Dynamics} (NAMD) software~\cite{phillips2005scalable}. 

 Our point of departure in this paper is a synergy of nonlinear KCM-based folding dynamics due to 
Kazerounian and collaborators~\cite{kazerounian2005protofold,tavousi2015protofold} and LQR 
control of folding due to Arkun and collaborators for \textcolor{black}{a linear time-invariant system}~\cite{arkun2010prediction,arkun2012combining}. 
In particular, we propose using a pointwise optimal
control law for guiding the protein folding simulations under the nonlinear 
KCM assumptions. Unlike the bead-spring LTI model in~\cite{arkun2010prediction,arkun2012combining} that uses 
the Cartesian coordinates of the atoms, 
the KCM model in~\cite{kazerounian2005protofold,tavousi2015protofold} directly accounts for the effect of the  
nonlinear interatomic interactions on the protein molecule dihedral angles. 
Our feedback control solution is based on the quadratic optimization of control torques, by using 
the optimal decision strategy (ODS) framework due to Barnard~\cite{barnard1975optimal}. Our QP-based control inputs minimize the deviation between the open-loop dynamics  
vector field and a reference model vector field under the entropy-loss constraints during the 
folding procedure. Optimal decision strategies 
have been employed in applications ranging from the control of manipulators with bounded 
inputs~\cite{spong1986control} and magnetic microrobots under 
frequency constraints~\cite{mohammadi2021integral} to controlling transients in power 
systems~\cite{thomas1976model}. ODS-based strategies belong to the larger family of optimization-based 
nonlinear controllers~\cite{murali2020graceful,klemm2020lqr,
	Ames17a,pandala2019qpswift}, whose applications in robotics are growing, thanks in part to advancements in mobile computation 
power.\\  
\noindent\textbf{Contributions of the paper.} This paper contributes to solving 
the protein conformation prediction and \textcolor{black}{obtaining more realistic folding pathways} 
using the KCM-based framework in 
two important ways. First, we demonstrate how the KCM-based protein folding 
difference equation can be cast as a control synthesis problem and that the kinetostatic torque 
in~\cite{kazerounian2005protofold,tavousi2015protofold} can be interpreted as 
a control input that guides the folding simulation (\textcolor{black}{Proposition~\ref{prop:trajRel}}). Second, 
using the control torque derived from the work in~\cite{kazerounian2005protofold,tavousi2015protofold}, we define a 
KCM-based reference vector field for the folding dynamics. This KCM-based vector field will then be used in 
a quadratic program (QP) for synthesizing bounded control inputs with predetermined bounds. Therefore, our 
synthesized control inputs will not only be close to the KCM-based reference 
vector field in~\cite{kazerounian2005protofold,tavousi2015protofold} but also bounded by user-specified bounds, thus  
respecting the requirement of avoiding high-entropy-loss routes during the folding process.   \textcolor{black}{Finally,  Proposition~\ref{prop:LipCont} provides sufficient conditions for uniqueness and 
	Lipschitz continuity of the control inputs generated by our proposed QP.}

The rest of this paper is organized as follows. First, we briefly review the kinematics and dynamics of 
protein folding in Section~\ref{sec:dynmodel}. Next, we demonstrate how the 
KCM-based approach to protein folding can be formulated as a control synthesis problem in  
Section~\ref{sec:ContProb}. Thereafter, we present 
our ODS-based control scheme for protein folding in Section~\ref{sec:Sol}. After 
presenting the simulation results in Section~\ref{sec:sims}, we 
conclude the paper with final remarks and 
future research directions in Section~\ref{sec:conc}.   

\noindent{\textbf{Notation.}} We let $\mathbb{R}_{+}$ denote the set of all non-negative 
real numbers.  Given a real number $x$, we let $\lfloor x \rfloor$ denote the greatest integer less than or equal to $x$.  Given $\mathbf{x}\in\mathbb{R}^M$, we let $|\mathbf{x}|:=\sqrt{\mathbf{x}^{\top}\mathbf{x}}$ and $|\mathbf{x}|_\infty:=\max\limits_{i} |x_i|$
denote the Euclidean and the maximum norms of $\mathbf{x}$, respectively. Given the integer $M$, we let $\mathbf{e}_M$ and $\mathbf{I}_M$ denote the vector $[1,\,\cdots,\,1]^\top$ and the identity matrix of size $M$, respectively. 

    \section{Preliminaries}
\label{sec:dynmodel} 
In this section, we briefly review the KCM framework for protein folding \emph{in 
vacuo}. For brevity, our presentation will be focused on the 
protein main chain. \textcolor{black}{This simplification is supported by the observation that the
motion of the backbone $\text{C}_\alpha$ 
atoms explains most of the essential folding
dynamics (see, e.g.,~\cite{amadei1993essential}). However, certain phenomenon such as hydrophobic core collapse can only be 
explained by considering the side chains~\cite{shao2010effects}.} Details on the inclusion of side chains can be found in~\cite{tavousi2015protofold,kazerounian2005protofold,kazerounian2005nano}. 
\subsection{Kinematic linkage representation of protein molecules}
\begin{figure}[t]
		\centering
		\includegraphics[width=0.48\textwidth]{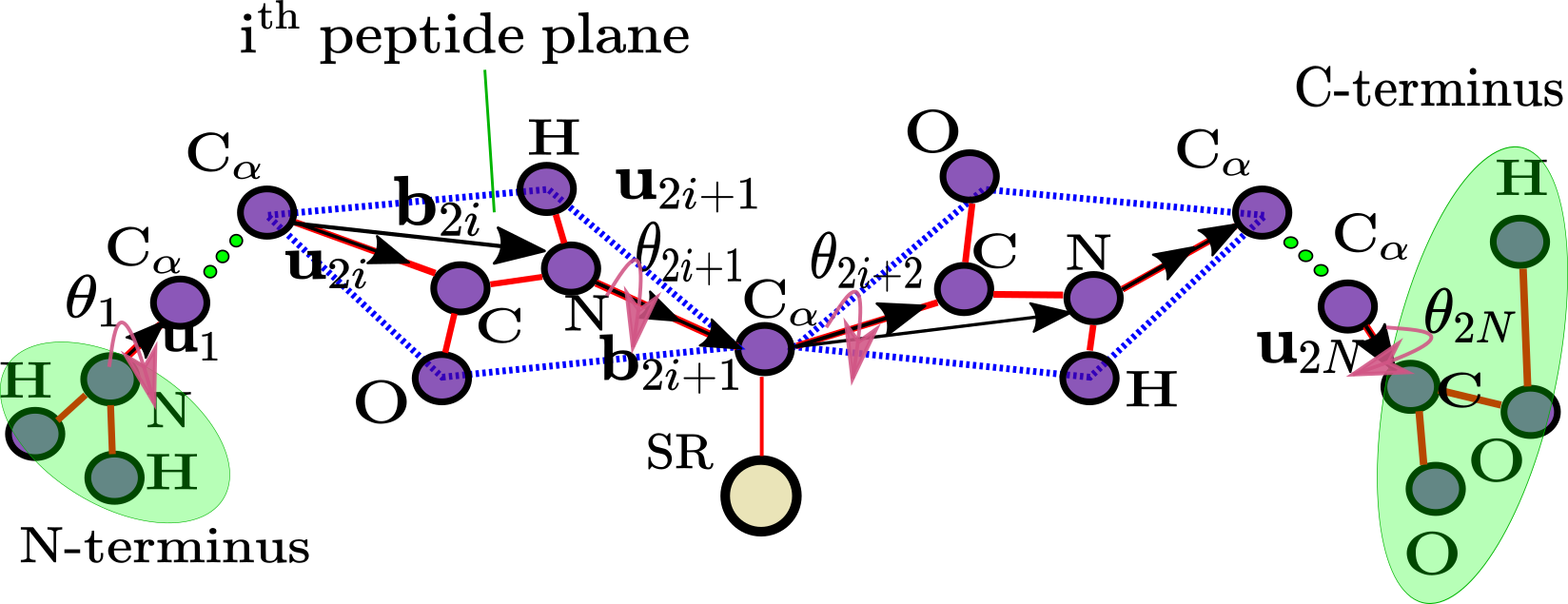}
		 \caption{\small The protein molecule kinematic structure under the 
		 	$\text{C}_\alpha-\text{CO}-\text{NH}-\text{C}_\alpha$ coplanarity assumption. For a 
	 	large-scaled version of this figure, see Supplementary Material.}
	 	\vspace{-1.5ex}
	 	\label{fig:protBasic}
\end{figure}

Proteins are long molecular chains consisting of peptide planes that are rigidly joined together 
via peptide bonds (see Figure~\ref{fig:protBasic}). Except the first and the last peptide planes,  
each pair of these planes are hinged to each other via a central carbon atom, known as the 
alpha-Carbon atom and denoted by $\text{C}_\alpha$. The coplanarity assumption of the 
six atoms $\text{C}_\alpha-\text{CO}-\text{NH}-\text{C}_\alpha$ in each amino acid is due to 
the experimental observations of the structure of protein molecules using high resolution X-ray crystallographic 
techniques~\cite{finkelstein2016protein}. In Figure~\ref{fig:protBasic}, the red line segments 
represent covalent chemical bonds between the atoms. A hydrogen atom is also covalently bonded to each $\text{C}_\alpha$ atom, 
which is not shown in the figure.
As it can be seen from Figure~\ref{fig:protBasic}, each 
$\text{C}_\alpha$ is connected to four other chemical
components, namely, the three atoms $\text{N}$, $\text{H}$, and $\text{C}$ 
as well as a variable side chain denoted by SR. The first $\text{C}_\alpha$ of the linkage is 
connected to an amino group, known as the N-terminus, and one other peptide plane.  The last $\text{C}_\alpha$ of 
the linkage hinges to a carboxyl group, known as the C-terminus, and one other peptide plane.  

The backbone conformation consisting of 
$-\text{N}-\text{C}_\alpha-\text{C}-$ atoms is completely determined  using a set of 
bond lengths and two sets of dihedral angles, i.e., the rotation angles around $\text{N}-\text{C}_\alpha$ and 
$\text{C}_\alpha-\text{C}$. Therefore, the vector 
$
\pmb{\theta}= \big[\theta_1,\cdots,\theta_{2N}]^\top \in \mathbb{R}^{2N}
$
represents the backbone chain configuration   with $N-1$ peptide planes. Corresponding to  
each DOF of the protein molecule, we consider a unit vector along its corresponding rotation axis and denote 
it by $\mathbf{u}_j$, $1\leq j \leq 2N$. Hence, the vectors $\mathbf{u}_{2i}$ and $\mathbf{u}_{2i+1}$ are the unit vectors 
along the $\text{C}_\alpha-\text{C}$ and $\text{N}-\text{C}_\alpha$ bonds of the $i$-th amino acid, respectively. Moreover, 
$\mathbf{u}_1$ and $\mathbf{u}_{2N}$ correspond to the unit vectors of the $\text{N}$- and $\text{C}$-termini, respectively.

In addition to the unit vectors $\mathbf{u}_j$, Kazerounian and collaborators 
use the so-called \textbf{body vectors} to complete the spatial orientation description 
of the rigid peptide planes. The body vectors are denoted by $\mathbf{b}_{j}$, $1\leq j \leq 2N$ 
and they completely determine the relative position of any two atoms in the peptide planes. In particular, the 
relative positions of any two atoms can be described by the linear combination $k_{1m} \mathbf{b}_{2i} + k_{2m} \mathbf{b}_{2i+1}$, 
where the constants $k_{1m}$ and $k_{2m}$, $1\leq m \leq 4$, are the same across all peptide planes (see~\cite{alvarado2003rotational,Tavousi2016}
for further details).   Using the vectors $\mathbf{u}_j$ and $\mathbf{b}_{j}$, one can completely describe the conformation of 
a protein molecule using the vector of dihedral angles $\pmb{\theta}$. In particular, after designating the zero position conformation with $\pmb{\theta}=\mathbf{0}$ corresponding to the biological reference position of the chain (see~\cite{kazerounian2005nano}), the transformations 
\begin{equation}
\mathbf{u}_j(\pmb{\theta}) = \Xi(\pmb{\theta},\mathbf{u}_j^0) \mathbf{u}_j^0  ,\, 
\mathbf{b}_j(\pmb{\theta}) = \Xi(\pmb{\theta},\mathbf{u}_j^0) \mathbf{b}_j^0,
\label{eq:kineProtu}
\end{equation}
where $\Xi(\pmb{\theta},\mathbf{u}_j^0)=\prod_{r=1}^{j} R(\theta_j,\mathbf{u}_j^0)$, can be used to describe the kinematic configuration of the protein molecule given 
the vector $\pmb{\theta}$. In~\eqref{eq:kineProtu}, the 
rotation matrix $R(\theta_j,\mathbf{u}_j^0)\in \text{SO}(3)$ represents the rotation about the unit vector 
$\mathbf{u}_j^0$ with angle $\theta_j$. Having obtained the body 
vectors $\mathbf{b}_j(\pmb{\theta})$ from~\eqref{eq:kineProtu} and under the assumption that the N-terminus atom is fixed at the 
origin, the coordinates of the backbone chain atoms in the $k$\textsuperscript{th}-peptide plane can be found from
$
\mathbf{r}_i(\pmb{\theta}) = \sum_{j=1}^{i}  \mathbf{b}_j(\pmb{\theta}), \;\; 1 \leq i \leq 2N - 1 
$
where the indices $i=2k-1$ and $i=2k$ correspond to the N and $\text{C}_\alpha$ atoms, respectively. 
\subsection{KCM-based folding}
The KCM framework in~\cite{kazerounian2005protofold,tavousi2015protofold,Tavousi2016} relies on the well-established fact that the inertial forces during the folding of a protein chain are negligible with respect to their electrostatic and van der Waals interatomic counterparts (see, e.g.,~\cite{adolf1991brownian,kazerounian2005protofold,arkun2010prediction,arkun2012combining}).  Instead, the dihedral angles change under the kinetostatic effect of the interatomic force fields with an amount proportional to the effective torques on these joints.

 To present the KCM framework, let us consider a protein chain with $N_a$ atoms and $N-1$ peptide planes with the dihedral angle conformation vector $\pmb{\theta}\in \mathbb{R}^{2N}$ and use the following notation. Given any two atoms $a_i$, $a_j$ in the 
 protein chain, we denote their Cartesian position vectors by $r_i(\pmb{\theta})$, $r_j(\pmb{\theta})$, their distance by $d_{ij}(\pmb{\theta}):= | r_i(\pmb{\theta})-r_j(\pmb{\theta})|$, their electrostatic charges by  $q_i$, $q_j$, their van der Waals radii by $R_i$, $R_j$, their van der Waals distance by $D_{ij}=R_i+R_j$, their dielectric constant by $\varepsilon_{ij}$, and their depth of potential well as $\epsilon_{ij}$. Finally, we let $w_{ij}^{\text{elec}}$ and $w_{ij}^{\text{vdw}}$  represent the weight factors for the electrostatic and van der Waals forces between the two atoms $a_i$ and $a_j$, respectively. All of these parameters are reported in~\cite{Tavousi2016} and the references therein.  
 
 The successive iteration of the KCM-based folding process is performed until all of the kinetostatic torques converge to a minimum corresponding to a local minimum of the aggregated free energy function of the protein chain. In particular, the aggregated free energy 
 is given by 
 \begin{equation}\label{eq:freePot}
 \mathcal{G}(\pmb{\theta}) := \mathcal{G}^{\text{elec}}(\pmb{\theta}) + \mathcal{G}^{\text{vdw}}(\pmb{\theta}),
 \end{equation}
 where 
 \begin{equation}\label{eq:elec}
 \mathcal{G}^{\text{elec}}(\pmb{\theta}) =  \sum_{i=1}^{N_\text{a}} \sum_{j\neq i} \frac{w_{ij}^{\text{elec}}}{4\pi \varepsilon_{ij}} \frac{q_i q_j}{d_{ij}(\pmb{\theta})}, 
 \end{equation}
 is the potential energy of the molecule due to the electrostatic interactions between the atoms, and 
 \begin{equation}\label{eq:vdw}
 \mathcal{G}^{\text{vdw}}(\pmb{\theta}) =  \sum_{i=1}^{N_\text{a}} \sum_{j\neq i} w_{ij}^{\text{vdw}}\epsilon_{ij} \bigg[\big(\frac{D_{ij}}{d_{ij}(\pmb{\theta})}\big)^{12}-2\big(\frac{D_{ij}}{d_{ij}(\pmb{\theta})}\big)^{6}\bigg]
 \end{equation}
 \noindent is the potential energy due to the van der Waals interactions.  The resultant Coulombic and Van der Waals forces on 
 each atom $a_i$, $1\leq i \leq N_\text{a}$, are given by $F_{i}^{\text{elec}}(\pmb{\theta})=-\nabla_{\mathbf{r}_i} \mathcal{G}^{\text{elec}}$ and $F_{i}^{\text{vdw}}(\pmb{\theta})=-\nabla_{\mathbf{r}_i}\mathcal{G}^{\text{vdw}}$, respectively.

 Computing the resultant forces and torques on each of the $N-1$ peptide planes, which are considered as the rigid links of the peptide 
 linkage, and appending them in the generalized force vector $\mathcal{F}(\pmb{\theta})\in \mathbb{R}^{6N}$, one can use a proper mapping 
 to relate $\mathcal{F}(\pmb{\theta})$ to the equivalent torque vector acting on the dihedral angles. 
 In particular, starting from an initial conformation vector 
 $\pmb{\theta}_0$, the KCM-based iteration can be written using the difference equation
\begin{equation}\label{eq:KCM_alg}
\pmb{\theta}_{k+1} =  \pmb{\theta}_{k} + \frac{h}{| \pmb{\tau}(\pmb{\theta}_k) |_\infty} \pmb{\tau}(\pmb{\theta}_k),\; 
\end{equation}
\noindent where $k$ is a non-negative integer and $h$ is a positive real constant that tunes the maximum dihedral angle rotation magnitude in each step. The vector $\pmb{\tau}(\pmb{\theta}_k)\in \mathbb{R}^{2N}$, which represents the overall joint torques due to the interatomic forces in the protein 
molecule, is given by 
\begin{equation}\label{eq:tau_KCM}
\pmb{\tau}(\pmb{\theta}_k) = \mathcal{J}^\top(\pmb{\theta}_k) \mathcal{F}(\pmb{\theta}_k).
\end{equation}
\noindent In~\eqref{eq:tau_KCM}, the matrix  $\mathcal{J}(\pmb{\theta}_k)\in \mathbb{R}^{6N\times 2N}$ 
represents the Jacobian of the protein chain at conformation $\pmb{\theta}=\pmb{\theta}_k$ (see~\cite{tavousi2015protofold,kazerounian2005protofold} for the 
calculation details). The vector  $\mathcal{F}(\pmb{\theta}_k)$  is due to the torques and forces acting on the peptide planes
at  $\pmb{\theta}=\pmb{\theta}_k$. 
   
    \section{KCM-based Folding Iteration as a Nonlinear Control-Affine System}
\label{sec:ContProb}

In this section we formulate the KCM-based folding in~\eqref{eq:KCM_alg},~\eqref{eq:tau_KCM} as a control 
synthesis problem. Such control theoretic point of view  
 can be traced back to the field of targeted molecular dynamics (TMD)~\cite{ferrara2000computer} 
(see also~\cite{arkun2011protein,arkun2012combining}) where the distance of the protein molecule 
to a target conformation is constrained by means of a proportional control input for accelerating 
the transition to the final structure. 

Consider the following nonlinear control-affine system
\begin{equation}\label{eq:KCM_ode}
\dot{\pmb{\theta}} = \mathcal{J}^\top(\pmb{\theta}) \mathcal{F}(\pmb{\theta}) + \mathbf{u}_{c},
\end{equation}
where the Jacobian matrix $\mathcal{J}(\pmb{\theta})$ and the generalized force vector $\mathcal{F}(\pmb{\theta})$ 
are the same as in~\eqref{eq:tau_KCM}. The nonlinear control-affine system in~\eqref{eq:KCM_ode} governs the evolution of the dihedral angles under the effect of interatomic forces and an additional control input for constraining the folding process simulation. 

\begin{remark}
In the optimal control synthesis considered by Arkun and collaborators~\cite{arkun2010prediction,arkun2011protein,arkun2012combining}, the atomic interactions are modeled by linear spring-like forces coupled via a proper connectivity matrix. \textcolor{black}{In this work, we are considering a nonlinear dynamical system (as opposed to the linear dynamics in~\cite{arkun2010prediction,arkun2011protein,arkun2012combining}) in the dihedral angle space of the molecule (as opposed to~\cite{arkun2010prediction,arkun2011protein,arkun2012combining}, where the state vector of the system is the Cartesian position vector of the $\text{C}_\alpha$ atoms). The Supplementary Material provides an in-depth comparison with~\cite{arkun2010prediction,arkun2011protein,arkun2012combining} (Section~B).}
\end{remark}

The following proposition relates the time trajectories of the nonlinear control-affine system in~\eqref{eq:KCM_ode} 
and the conformation samples obtained from the iterations in~\eqref{eq:KCM_alg} and~\eqref{eq:tau_KCM}. 
\begin{proposition}\label{prop:trajRel}
Consider the KCM-based iteration in~\eqref{eq:KCM_alg},~\eqref{eq:tau_KCM}, the nonlinear control-affine system in~\eqref{eq:KCM_ode}, and the state feedback control law
\begin{equation}\label{eq:KCM_control}
\mathbf{u}_{c}:= \mathbf{u}_{\text{KCM}}(\pmb{\theta}) = \frac{1-|\mathcal{J}^\top(\pmb{\theta}) \mathcal{F}(\pmb{\theta})|_\infty}{|\mathcal{J}^\top(\pmb{\theta}) \mathcal{F}(\pmb{\theta})|_\infty} \mathcal{J}^\top(\pmb{\theta}) \mathcal{F}(\pmb{\theta}).
\end{equation}	
Given an initial conformation $\pmb{\theta}=\pmb{\theta}_0$ and a final time $t^{\ast}>0$,  assume that $\sup \big|\mathcal{J}^\top(\pmb{\theta}) \mathcal{F}(\pmb{\theta})\big| \leq \lambda$ for some $\lambda>0$ in a neighborhood $\mathcal{N}_{\pmb{\theta}_0}$ of $\pmb{\theta}_0$. Moreover, suppose that the Lipschitz condition $\big|\mathcal{J}^\top(\pmb{\theta}) \mathcal{F}(\pmb{\theta})-\mathcal{J}^\top(\pmb{\theta}^\prime) \mathcal{F}(\pmb{\theta}^\prime)\big|\leq \lambda^{\prime} | \pmb{\theta}^\prime- \pmb{\theta}|$ holds for some $\lambda^{\prime}>0$, and all $\pmb{\theta}$, $\pmb{\theta}^\prime$  in the neighborhood $\mathcal{N}_{\pmb{\theta}_0}$. Then, the trajectory  of~\eqref{eq:KCM_ode} starting from $\pmb{\theta}_0$ under the control law in~\eqref{eq:KCM_control}, as long as it remains in  $\mathcal{N}_{\pmb{\theta}_0}$ during the time interval $[0,\, t^{\ast}]$, are related to the solutions of the KCM-based difference equation~\eqref{eq:KCM_alg},~\eqref{eq:tau_KCM} through
\begin{eqnarray}
\nonumber
\big| \pmb{\theta}(hk) - \pmb{\theta}_k \big| \leq \frac{1+\lambda}{2}(\exp(t^{\ast}\lambda^\prime)-1) h,\; \\
 \text{ for all } k = 0,\, 1,\, \cdots,\, \lfloor \frac{t^\ast}{h} \rfloor
\label{eq:KCM_diff_ode_relation}
\end{eqnarray}
\end{proposition}
\begin{proof} 
	\textcolor{black}{See Section D in Supplementary Material.}  
\end{proof}

\begin{remark}
Proposition~\ref{prop:trajRel} implies that the tuning parameter $h$  in ~\eqref{eq:KCM_alg} can be interpreted as the 
sampling time step for a forward Euler discretization of~\eqref{eq:KCM_ode} under the control input~\eqref{eq:KCM_control}.  Although this control interpretation is not present in the work of Kazerounian and collaborators~\cite{kazerounian2005protofold,Tavousi2016,tavousi2015protofold}, it will enable us to extend the KCM framework and obtain more realistic folding pathways under the KCM-based simulations. 
\end{remark}

Using the KCM-based control input~\eqref{eq:KCM_control} in~\eqref{eq:KCM_ode}, we arrive at the following reference vector field
\begin{equation}\label{eq:KCM_vecField}
\dot{\pmb{\theta}}^{\text{KCM}} := \frac{\mathcal{J}^\top(\pmb{\theta}) \mathcal{F}(\pmb{\theta})}{|\mathcal{J}^\top(\pmb{\theta}) \mathcal{F}(\pmb{\theta})|_\infty}.
\end{equation}
The prominent KCM-based schemes by Kazerounian and collaborators~\cite{kazerounian2005protofold,Tavousi2016,tavousi2015protofold} for folding \emph{in vacuo} are therefore governed by the above reference vector field in the protein conformation landscape, according to the forward Euler iterations given by~\eqref{eq:KCM_alg} and~\eqref{eq:tau_KCM}.  
    
     \section{ODS-Based Control of Protein Folding}
\label{sec:Sol}

In this section, we will use the KCM-based reference vector field in~\eqref{eq:KCM_vecField} for synthesizing QP-based control inputs for the folding dynamics in~\eqref{eq:KCM_ode}. The control synthesis interpretation of protein folding enables us to add more constraints to the folding simulation and hence making it closer to the true folding structural biochemistry.\\
\indent One important physical observation  in the structural biochemistry of folding is that to avoid extremely fast decay of excess entropy, one needs to penalize the control input during the guided folding process (see~\cite{arkun2010prediction,arkun2011protein,weikl2003folding}). Therefore, the dihedral angle vector trajectories should not only follow the KCM-based reference vector field in~\eqref{eq:KCM_vecField} but also the control torques need to be bounded by a predetermined bound throughout the simulation.  \textcolor{black}{As it is shown in ``Entropy-Loss Rate of the Protein Molecule'' (Section A) in Supplementary Material, the rate of change of the protein molecule entropy linearly depends on the control input $\mathbf{u}_c$ in~\eqref{eq:KCM_ode}. Furthermore, the upper bound on this rate of change depends on the largest element of the control input vector.}\\
\indent ODS-based control is a pointwise optimal control scheme
that, in the context of protein folding, minimizes the deviation between the closed-loop vector field of 
the protein molecule dynamics in~\eqref{eq:KCM_ode} and the KCM-based reference vector field in~\eqref{eq:KCM_vecField}, while 
respecting the input constraints 
\begin{equation}\label{eq:cont_bounds}
|u_{c_{_i}}| \leq c_i, \; i = 1,\, \cdots,\, 2N, 
\end{equation}
where $u_{c_{_i}}$ is the $i$\textsuperscript{th} control input.

\begin{figure}[t]
	\vspace{-2pt}
	\centering
	\includegraphics[width=0.3\textwidth]{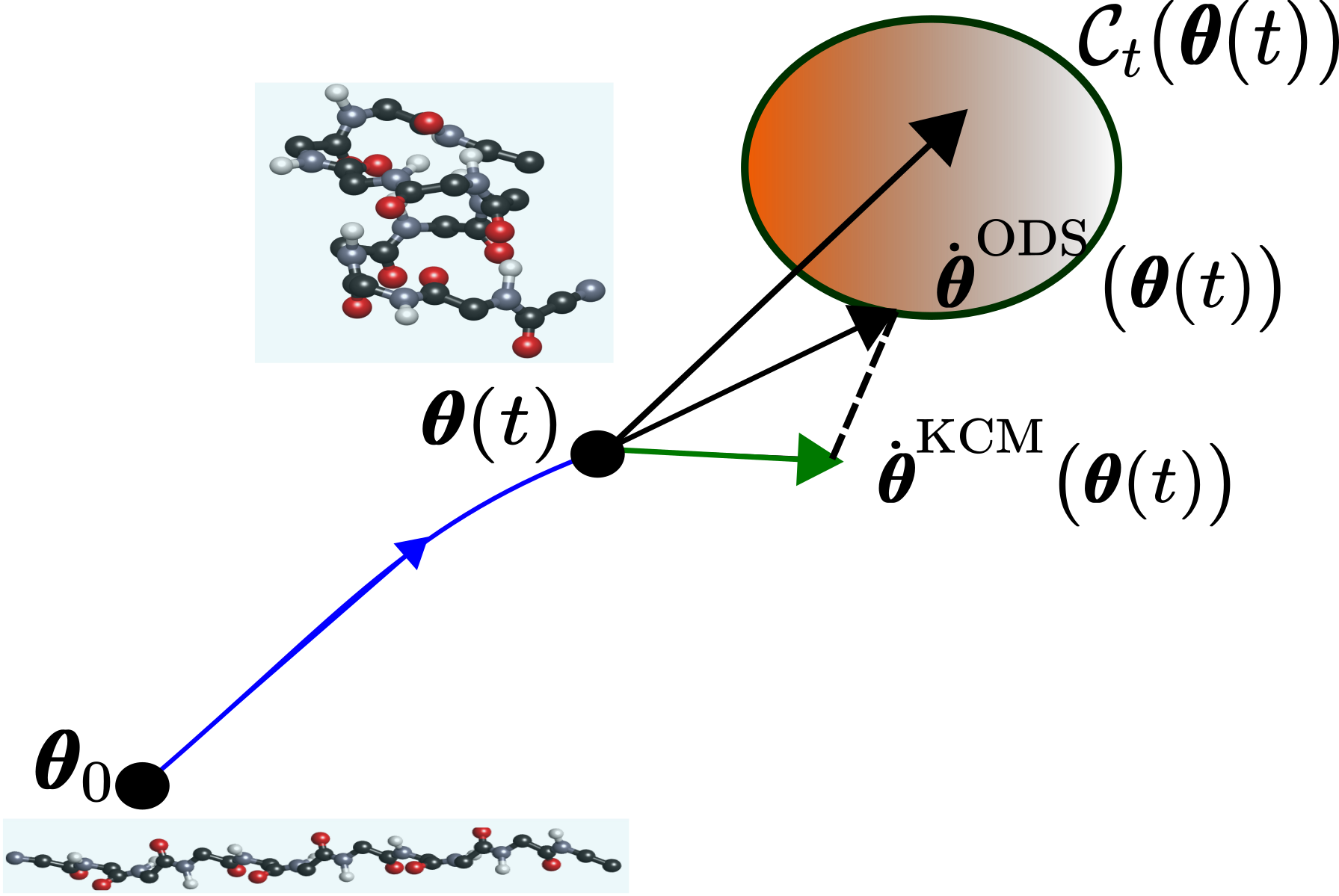}
	\caption{\small Optimal decision strategy in the context of protein folding.}
	\vspace{-2ex}
	\label{fige:ODS_Geom}
\end{figure}

In order to state the ODS-based control scheme 
for KCM-based protein folding, let $\pmb{\theta}(t,t_0,\pmb{\theta}_0,\mathbf{u}_c(t))$, or 
$\pmb{\theta}(t)$ for short, denote the solution to~\eqref{eq:KCM_ode} under the control input 
$t\mapsto \mathbf{u}_c(t)$ and initial conformation $\pmb{\theta}_0$ at time $t_0$. For each solution $\pmb{\theta}(t)$, we 
define the set of permissible velocity vectors $\mathcal{C}_t(\pmb{\theta}(t))$ to be the translation by $\pmb{\theta}(t)$ of the set 
of \textcolor{black}{free vectors}
\begin{equation}\label{eq:permissible}
\begin{aligned}
\mathcal{C}(\pmb{\theta}(t)) := \big\{ \mathbf{v}(t,\mathbf{u}_{c}) \in \mathbb{R}^{2N} \big| \mathbf{v} = \mathcal{J}^\top(\pmb{\theta}) \mathcal{F}(\pmb{\theta}) +  \mathbf{u}_{c},\, \\
|\mathbf{u}_{c}{_{_i}} |\leq c_i \big\}. 
\end{aligned}
\end{equation}
\textcolor{black}{\vspace{-1pt}In other words, each vector in $\mathcal{C}_t(\pmb{\theta}(t))$ corresponds to a 
	vector in $\mathcal{C}(\pmb{\theta}(t))$ whose initial point has been translated to $\pmb{\theta}(t)$.} Therefore, for any control input $\mathbf{u}_c(t)$ that respects the input constraints in~\eqref{eq:cont_bounds}, 
the possible tangent vectors to the folding pathway lie in $\mathcal{C}_t(\pmb{\theta}(t))$ 
(see Figure~\ref{fige:ODS_Geom}).  The protein conformation  
evolves according to the closed-loop vector field $\dot{\pmb{\theta}}^{\text{ODS}}$ that is not only close to the KCM-based reference vector field $\dot{\pmb{\theta}}^{\text{KCM}}$ in~\eqref{eq:KCM_vecField} but also satisfies the control input constraints in~\eqref{eq:cont_bounds}.
In the context of protein folding, we choose the ODS-based control law $\mathbf{u}_c(t)$ in a way that at each time $t$, the instantaneous 
closed-loop vector field is the ``nearest'' to the KCM-based reference vector field $\dot{\pmb{\theta}}^{\text{KCM}}(\pmb{\theta}(t))$  
given by~\eqref{eq:KCM_vecField} in the norm on $\mathbb{R}^{2N}$ defined by some positive definite matrix $\mathbf{Q}$. In other words, the control law at each $t$ is the minimizing solution to the QP 
\begin{equation}\label{eq:permissible2}
\begin{aligned}
\underset{\mathbf{v}\in \mathcal{C}_t(\pmb{\theta}(t))}{\text{min.}} \;  \Big\{  \big[\mathbf{v} - \dot{\pmb{\theta}}^{\text{KCM}}(\pmb{\theta}(t)) \big]^{\top} \mathbf{Q}   \big[ \mathbf{v} - \dot{\pmb{\theta}}^{\text{KCM}}(\pmb{\theta}(t)) \big] \Big\},  
\end{aligned}
\end{equation}
%
%
which can be shown to be equivalent to the QP
\begin{equation}\label{eq:ODS}
\begin{aligned}
\mathbf{u}_c^{\ast}(\pmb{\theta}) = \underset{\vx{u}}{\text{argmin}} \; & \big\{ \frac{1}{2} \vx{u}_c^{\top} \mathbf{Q} \vx{u}_c + \vx{G}^{\top}(\pmb{\theta}) \vx{u}_c \big\} \\
\text{subject to} & \;\;\; |\mathbf{u}_{c}{_{_i}} |\leq c_i, 
\end{aligned}
\end{equation}
where
\begin{equation}\label{eq:ODS2}
\textcolor{black}{
	\begin{aligned}
	\vx{G}(\pmb{\theta}) := - \mathbf{Q} \big( \mathcal{J}^\top(\pmb{\theta}) \mathcal{F}(\pmb{\theta})  -  \dot{\pmb{\theta}}^{\text{KCM}}(\pmb{\theta}) \big). 
	\end{aligned}
}
\end{equation}
\textcolor{black}{ The following proposition provides sufficient conditions for uniqueness and Lipschitz continuity of the 
	proposed QP in~\eqref{eq:ODS} at each conformation $\pmb{\theta}_0$. 
\begin{proposition}\label{prop:LipCont}
	Consider the QP~\eqref{eq:ODS}, and assume the
following conditions at an 
	arbitrary conformation $\pmb{\theta}_0$:
\begin{itemize}
	\item The unique solution to the linear programming problem 
	\begin{equation}\label{eq:unique}
	\begin{aligned}
	\omega = \underset{(\mathbf{u}_c,w)\in \mathbb{R}^{2N+1}}{\text{max}} \; & w \\
	\text{subject to} & \;\;\; \mathbf{A} \mathbf{u}_\text{c} + w \mathbf{e}_{_{4N}}  \leq \mathbf{c}, 
	\end{aligned}
	\end{equation}
	satisfies $\omega>0$, where the vector $\mathbf{c}\in \mathbb{R}^{4N}$ is determined by the constant bounds  
	$c_i$ in~\eqref{eq:cont_bounds} as $\mathbf{c}=[c_1,\,c_1,\, \cdots,\, c_{2N},\, c_{2N}]^{\top}$; moreover, the 
	matrix $\mathbf{A}\in \mathbb{R}^{4N\times 2N}$ is 
	given by 
	$
	\begin{psmallmatrix}
	   1 & 0 & \cdots & & 0 \\
	  -1 & 0 & \cdots & & 0 \\
	  \vdots & \vdots & & & \vdots\\
	  0 &    &        & & 1\\
	  0 &    &        & & -1
	\end{psmallmatrix} ,
	$ 
	\item $\mathbf{Q}$ in~\eqref{eq:ODS} is symmetric and positive definite,
	\item $ \mathcal{J}^\top(\pmb{\theta}) \mathcal{F}(\pmb{\theta})$ is Lipschitz continuous at $\pmb{\theta}_0$.
\end{itemize}
	Then, the feedback $\mathbf{u}_c^{\ast}(\pmb{\theta})$ defined in~\eqref{eq:ODS} is Lipschitz continuous and 
	unique at conformation $\pmb{\theta}_0$. 
\end{proposition}
\begin{proof}
See Section C in Supplementary Material. 
\end{proof}
}
\textcolor{black}{
	\vspace{-2ex}
\begin{remark}
	When the constraints on the control inputs in~~\eqref{eq:cont_bounds} are sufficiently large, or in the limit when $c_i\to \infty$, the generated control 
	inputs by the proposed QP in~\eqref{eq:ODS} will be the same as the conventional KCM control torques in~\eqref{eq:KCM_control}.
\end{remark}
}
     
     \section{Simulation Results}
\label{sec:sims}

In this section we present numerical simulation results to validate 
our proposed control method. We consider a molecular chain with $N-1=10$ peptide planes 
corresponding to a $2N = 22$-dimensional dihedral angle space in~\eqref{eq:KCM_ode}. All of 
our implementation has been done in MATLAB \textcolor{black}{R2018b} following the PROTOFOLD~I
guidelines~\cite{kazerounian2005protofold,Tavousi2016} \textcolor{black}{on an Intel\textsuperscript{\textregistered} Core\textsuperscript{\texttrademark} i7-6770HQ CPU@2.60GHz}. In our current implementation, the side chains are ignored. \textcolor{black}{See Section E of Supplementary Material for the implementation  details and computational complexity analysis of the overall algorithm.} 

\begin{figure*}
	\vspace{-1.5ex}
	\centering
    \begin{subfigure}{0.35\textwidth}
	\centering
	\includegraphics[width=\textwidth]{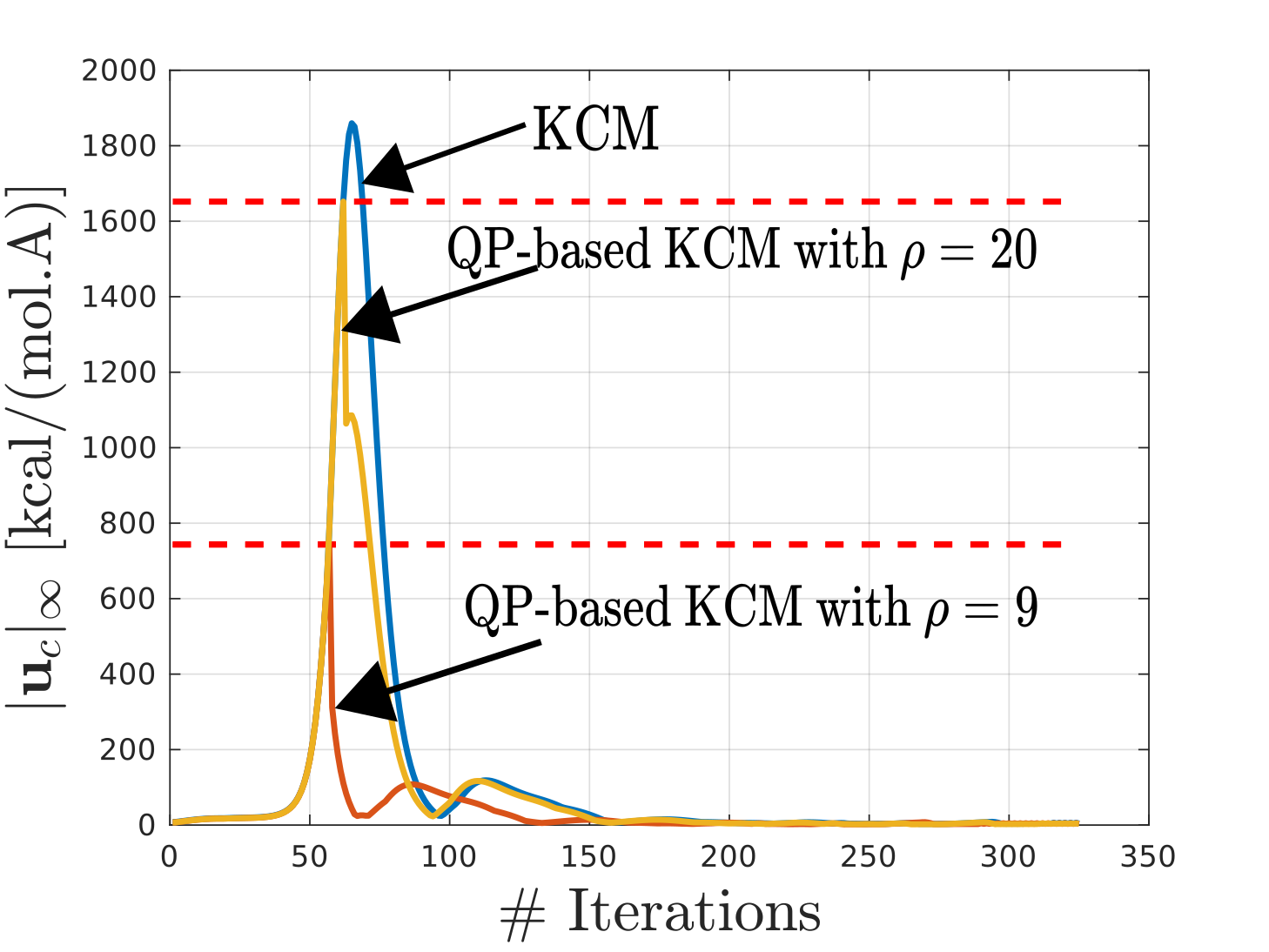} 
	\caption{Joint torque profiles during protein folding.}
	\label{fig:jointTorques}
\end{subfigure}
	\begin{subfigure}{0.35\textwidth}
		\vspace{-0.5ex}
		\centering
		\includegraphics[width=\textwidth]{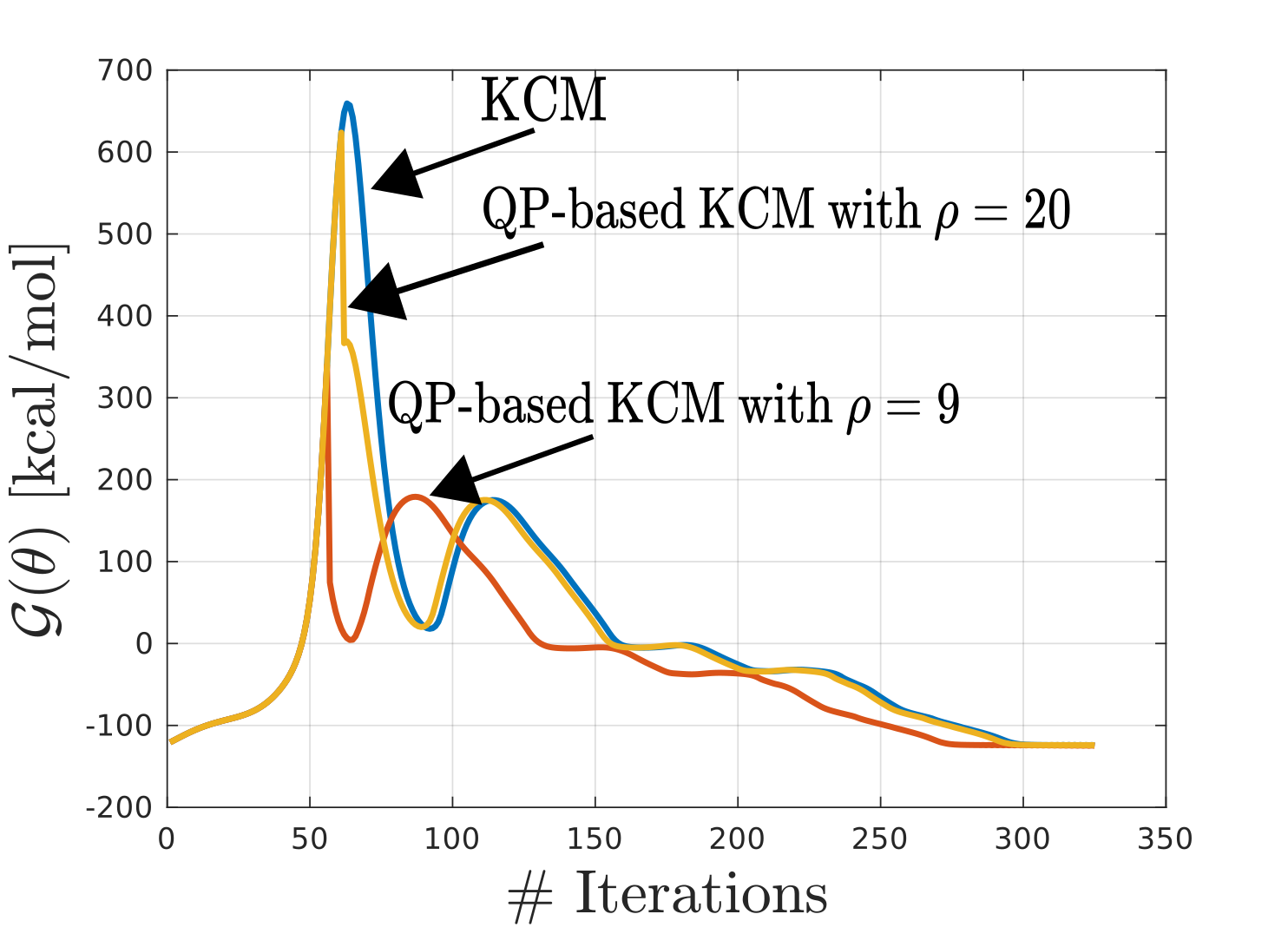} 
		\caption{Energy profiles during protein folding.}
		\label{fig:freeEnergy}
	\end{subfigure}
\begin{subfigure}{0.20\textwidth}
	\centering
	\includegraphics[width=\textwidth]{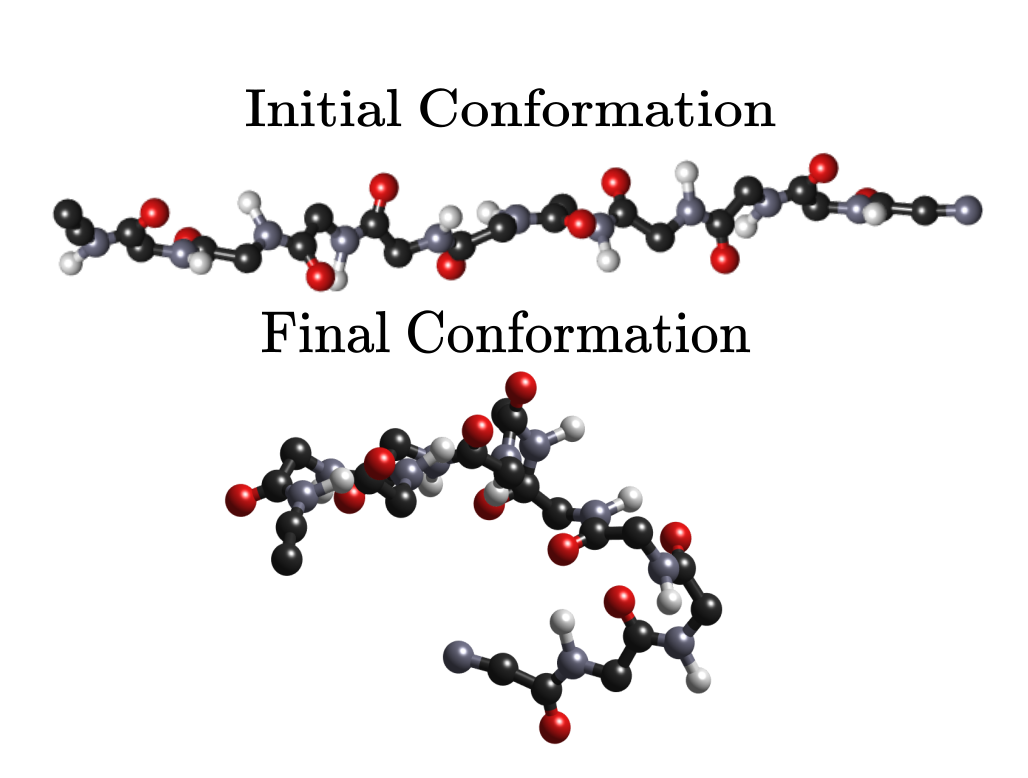} 
	\vspace{3ex}
	\caption{Initial and final conformations.}
	\label{fig:initFinal}
\end{subfigure}
\caption{Simulation results using the conventional and the QP-based KCM control torques with two different sets of constraints.}
\end{figure*}

We run three sets of simulations. \textcolor{black}{Our simulations are based on the forward 
	Euler discretization of the underlying continuous time dynamics in~\eqref{eq:KCM_ode}. Propositions~\ref{prop:trajRel} and~\ref{prop:LipCont} guarantee the ``well-behavedness'' of the resulting trajectories under small enough time steps and not too tight constraints on the control inputs.}  The first simulation is run by using the original KCM-based control torque in~\eqref{eq:KCM_control} with 
$h=0.04$ and for \textcolor{black}{325 iterations starting from  a vector of initial dihedral angles 
uniformly distributed about the mean value $27.7^{^\circ}$ with standard deviation $1.1^{^\circ}$\footnote{See Supplementary Material 
	for the exact value of $\pmb{\theta}_0$.  
}. The time it takes to run the program is equal to 3.47 \textsuperscript{sec}.}
	As expected from the KCM-based folding scheme~\cite{kazerounian2005protofold,tavousi2015protofold}, the 
torques converge to a small neighborhood of zero and the free energy of the molecule also converge to a local minimum on the folding landscape. The blue curves in~\ref{fig:jointTorques} and~\ref{fig:freeEnergy} depict the profiles of the maximum control torque $|\mathbf{u}_c|_\infty$ and the aggregated free energy $\mathcal{G}(\pmb{\theta})$ \textcolor{black}{of the conventional KCM framework} versus the number of iterations, respectively. Next, we constrain the control torques used for guiding the folding process according to the QP-based control scheme in~\eqref{eq:ODS} while using the KCM reference vector field in~\eqref{eq:KCM_vecField}. We use 
\begin{eqnarray}\label{eq:Q_c}
\mathbf{Q} = \sqrt{2N} \mathbf{I}_{2N},\; c_i = c_0 \frac{\rho}{\sqrt{2N}},\, 1 \leq i \leq 2N,
\end{eqnarray}
for the quadratic weight matrix and the constraints on the folding control torques, respectively. In~\eqref{eq:Q_c}, the constant factor $c_0$ for conversion to $\text{kcal}/\text{mol}$ units of energy is equal to $N_0 q_e^2 k_{\text{J-kcal}}\times 10^{20}$ where $N_0=6.022 \times 10^{23}$ is the Avogadro number, $q_e = 1.602 \times 10^{-19}$ is the charge of 
electron in coulombs, $k_{\text{J-kcal}}=1/4184$ is the energy unit conversion factor from Joules to kilo calories. The parameter $\rho$ in~\eqref{eq:Q_c} is used for relaxing/tightening the constraints on the control torques. 
The orange (associated with $\rho=20$) and red (associated with $\rho=9$) curves in~\ref{fig:jointTorques} and~\ref{fig:freeEnergy} depict the profiles of $|\mathbf{u}_c|_\infty$ and the free energy of the peptide chain $\mathcal{G}(\pmb{\theta})$ versus the number of iterations, respectively. As it can be seen from~\ref{fig:jointTorques}, the QP-based control torques respect the predetermined bounds while both converging to zero. \textcolor{black}{It takes around 4.1\textsuperscript{sec} to run 
	each of these two simulations. This should not be surprising as we are running a QP to respect the control input constraints 
during folding (see Section E of Supplementary Material). Indeed, the increased computational 
complexity is the price to be paid for obtaining more realistic folding pathways.}   \textcolor{black}{Next, to violate the first condition in Proposition 4.1, we gradually made the constraints tighter to the point of failure at around $\rho=2$, where QP-based control fails to exist at certain conformations along the folding pathway.}  \textcolor{black}{While both KCM and QP-based KCM converge to the same local minimum of the protein molecule energy, the folding pathways 
generated by the QP-based KCM result in both smaller control torques and transient energy values; hence, respecting the biological observations about the entropy-loss constraints during folding.}\vspace{-1pt}

     \section{Concluding Remarks and Future Research Directions}
\label{sec:conc}
\textcolor{black}{Considering the importance of computer-aided prediction of the folded structure of 
	protein molecules and their folding pathways, this paper formally extended the KCM-based framework for protein conformation 
	prediction by including entropy-loss constraints via a QP-based control solution.  Indeed, constraining the 
	entropy-loss rates during the folding simulations would make the obtained pathways more realistic. After relating the conventional KCM iterations to a proper nonlinear affine control system, we demonstrated that the well-established KCM torques can be utilized for defining a reference vector field for QP-based control schemes that account for the entropy-loss constraints on the folding pathway. Furthermore, we provided sufficient conditions for uniqueness and Lipschitz continuity of the generated control torques at each conformation of the protein molecule.  Simulations 
for a simplified molecular chain demonstrated the effectiveness of the proposed methodology in obtaining more realistic 
folding pathways. These findings lead us to further research avenues such as investigating the folding pathways for 
two-state proteins, protein misfolding dynamics, and developing adaptive schemes for TMD-based folding.}



\vspace{-1ex}
\bibliographystyle{./styles/IEEEtran}
\bibliography{PhDBiblio}
\newpage

\newcounter{defcounter}
\setcounter{defcounter}{0}
\newenvironment{myequation}{%
	\addtocounter{equation}{-1}
	\refstepcounter{defcounter}
	\renewcommand\theequation{s\thedefcounter}
	\begin{equation}}
{\end{equation}}
\renewcommand\thefigure{SF\arabic{figure}}
\setcounter{figure}{0}   
\section*{Supplementary Material To ``Quadratic Optimization-Based Nonlinear Control for Protein Conformation 
	Prediction''}

\noindent{\textbf{Notation.}} In these supplementary notes, we will number the equations using (s\#), the references using [S\#], and the figures using SF\#. Therefore, (s1) refers to Equation~1 in the supplementary notes while (1) refers to Equation~1 in the original article. The same holds true for the references and the figures. 

\subsection{Entropy-Loss Rate of the Protein Molecule}
There are a few equivalent ways to compute the entropy changes of a protein molecule during folding including the 
methods based on configurational entropy computations and the methods based on thermodynamics-based operational 
definitions. Here, we will use Clausius thermodynamics expression in [S1], which is similar to  
the derivations in~\cite{arkun2010prediction} in that it is based on classical thermodynamics arguments.   
Based on Clausius equation, the change of the molecule 
entropy is given by 
\begin{myequation}
\Delta S(\pmb{\theta})  = \kappa_0 \Delta \mathcal{G}(\pmb{\theta}),  
\end{myequation}
\hspace{-1.5ex} where $\kappa_0$ is a temperature-dependent constant and $\mathcal{G}(\cdot)$ is the 
aggregated free energy of the molecule given by~\eqref{eq:freePot}. Hence, 
\begin{myequation}
\frac{d}{dt} S(\pmb{\theta})  = \kappa_0 \frac{\partial \mathcal{G}}{\partial \pmb{\theta}} \dot{\pmb{\theta}} = 
\kappa_0 \frac{\partial \mathcal{G}}{\partial \pmb{\theta}} \big( 
\mathcal{J}^\top(\pmb{\theta}) \mathcal{F}(\pmb{\theta}) + \mathbf{u}_{c}\big). 
\label{eq:ineqS}
\end{myequation}
\hspace{-1.5ex} Using H\"{o}lder's and the triangle inequalities in~\eqref{eq:ineqS}, it can be seen that
\begin{myequation}
	|\dot S|  \leq  \kappa_0 \bigg|\frac{\partial \mathcal{G}}{\partial \pmb{\theta}}\bigg|_1  \bigg( 
	\big|\mathcal{J}^\top(\pmb{\theta}) \mathcal{F}(\pmb{\theta})\big|_\infty + \big|\mathbf{u}_{c}\big|_\infty\bigg). 
\end{myequation}
\hspace{-1.5ex} where $|\cdot|_1$ denotes the vector 1-norm.

\subsection{In-Depth Comparison with the Optimal Control Framework in~\cite{arkun2010prediction,arkun2011protein,arkun2012combining}}

One of the inspirations for our ODS-based control scheme for protein folding in~\eqref{eq:ODS} has been due to the pioneering work 
by Arkun and collaborators~\cite{arkun2010prediction,arkun2011protein,arkun2012combining}. Here, we provide an in-depth comparison with our work and state possible extensions of the work by Arkun and collaborators~\cite{arkun2010prediction,arkun2011protein,arkun2012combining} to our nonlinear setting.  
\begin{itemize}
	\item \textbf{Difference in the studied dynamics:} As we stated in Remark 3.1 of the article, we are considering a \emph{nonlinear dynamics} (as opposed to the \emph{linear dynamics} in~\cite{arkun2010prediction,arkun2011protein,arkun2012combining}) in the dihedral angle space of the molecule (as opposed to~\cite{arkun2010prediction,arkun2011protein,arkun2012combining}, where the state vector of the system is the Cartesian position of the $\text{C}_\alpha$ atoms). 
	
	\item \textbf{Nonlinear extension of~\cite{arkun2010prediction,arkun2011protein,arkun2012combining}:} In the linear time-invariant (LTI) setup considered by Arkun and collaborators~\cite{arkun2010prediction,arkun2011protein,arkun2012combining}, the control input is penalized  by using an infinite horizon LQR cost function to avoid extremely fast decay of excess entropy. In the context of the 
	KCM framework, a cost function in the dihedral angle space can be written similar to~\cite{arkun2010prediction,arkun2011protein,arkun2012combining} as 
	\begin{myequation}\label{eq:ArkunCost}
	J[\mathbf{u}_c] := \int_{0}^{T_f} \big( \tilde{\pmb{\theta}}(t)^\top \mathbf{Q} \tilde{\pmb{\theta}}(t)+ \rho \mathbf{u}_c(t)^\top \mathbf{P} \mathbf{u}_c(t) \big) dt, 
	\end{myequation}
	\hspace{-1.5 ex} where $\tilde{\pmb{\theta}}:= \pmb{\theta} - \pmb{\theta}^{\text{ref}}$ is the error from the current to a reference 
	conformation, e.g., the native conformation of the protein. Furthermore, the tunable parameter $\rho>0$ is employed to make a trade-off 
	between avoiding the high
	energy regions of the landscape while choosing entropically preferred 
	folding pathways by penalizing high-entropy-loss routes. \underline{The similarities with~\cite{arkun2010prediction,arkun2011protein,arkun2012combining} stop here.}  Under an LTI control setup (which is not the case for the nonlinear dynamics in~\eqref{eq:KCM_ode}) and letting $T_f \to \infty$, the solution to~\eqref{eq:ArkunCost} takes the form of the feedback control law $\mathbf{u}_c^\ast(\pmb{\theta})=-\mathbf{K}(\rho)\tilde{\pmb{\theta}}$, where the state 
	feedback gain matrix $\mathbf{K}(\rho)$ is the solution to a proper algebraic Riccati equation (ARE).  However, this ARE-based feedback control law
	is not the minimzer of the cost function in~\eqref{eq:ArkunCost} when the dynamics are governed by~\eqref{eq:KCM_ode}. \\
	\hspace{2pt} Indeed, if the system has nonlinearities, which is the case in this article, there is a need for solving a Hamilton-Jacobi equation (HJE). HJEs are partial differential equations whose explicit closed-form solutions do not exist, even in the case of a simple
	nonlinearity (see, e.g., [S3]).  \\
	\hspace{3ex} Hamilton-Jacobi equation (HJE) for the protein folding dynamics given by~\eqref{eq:KCM_ode} can obtained as follows. Let us consider the nonlinear dynamics in~\eqref{eq:KCM_ode} and the cost function in~\eqref{eq:ArkunCost}. Defining $\tilde{\pmb{\theta}}:=\pmb{\theta} - \pmb{\theta}^{\text{ref}}$,  $\tilde{\mathbf{G}}(\tilde{\pmb{\theta}}):=\mathcal{J}^\top(\tilde{\pmb{\theta}}+\pmb{\theta}^{\text{ref}})\mathcal{F}(\tilde{\pmb{\theta}}+\pmb{\theta}^{\text{ref}})$, the 
	optimal control input is given by [S4]
	\begin{myequation}
	\mathbf{u}_c^{\ast}(\tilde{\pmb{\theta}}) = -\frac{1}{2\rho} \mathbf{P}^{-1} \nabla V(\tilde{\pmb{\theta}})+
	\tilde{\mathbf{G}}(\mathbf{0}). 
	\label{eq:optimalCont}
	\end{myequation}
	\hspace{-1.5 ex} where $V(\cdot)$ is a twice continuously differentiable function that satisfies HJE 
	\begin{myequation}
	\frac{1}{4\rho} \nabla V^\top \mathbf{P}^{-1} \nabla V - ( \mathcal{J}^\top(\pmb{\theta}) \mathcal{F}(\pmb{\theta}) - \tilde{\mathbf{G}}(\mathbf{0}) ) \nabla V - \tilde{\pmb{\theta}}^\top \mathbf{Q} \tilde{\pmb{\theta}} = 0.
	\label{eq:HJE}
    \end{myequation}
	For solving the above numerically demanding partial differential equation, there are a few methods such as employing artificial
	neural networks (see, e.g., [S4]). 
	
    \item \textbf{Computational load:} Even if we neglect the complexity arising from solving HJEs in a nonlinear setting, the framework in~\cite{arkun2010prediction,arkun2011protein,arkun2012combining} requires \emph{at least twice the computational time} of our framework. This is due to the fact that the framework in~\cite{arkun2010prediction,arkun2011protein,arkun2012combining} requires knowing the final folded structure of the protein to find the proper folding pathways. Our work, on the other hand, does not require knowing this final folded structure. Rather, by following the reference vector field coming from the KCM framework, folding starts from a random structure and converges to a local minimum of the aggregate free energy of the molecule. Therefore, a fair comparison with the nonlinear extension of~\cite{arkun2010prediction,arkun2011protein,arkun2012combining} would be to run the simulations in the dihedral angle space twice; namely, finding the final folded structure $\pmb{\theta}^{\text{ref}}$ using the conventional KCM simulation in the first run, and, in the second run, using the nonlinear extension of the optimal control framework in~\cite{arkun2010prediction,arkun2011protein,arkun2012combining} (as formulated in~\eqref{eq:optimalCont},~\eqref{eq:HJE}) to guide the molecule towards the final conformation $\pmb{\theta}^{\text{ref}}$, which has been obtained from the first run.
    
    \item \textbf{The underlying philosophy of the pointwise optimal control laws:}  ODS-based  control and its resulting QP-based synthesis~\cite{spong1986control} were originally developed to address some of the challenges arising from integral performance measures such as the one in~\eqref{eq:ArkunCost}. Indeed, the QP-based synthesis avoids problems such as computationally intensive schemes and  bypassing the Pontryagin maximum principle and/or HJEs under the bound constraints on the
    allowable input torques. Even ignoring the computational complexity of such schemes, one might prefer an instantaneous or pointwise optimization procedure which optimizes the
    present state of the system without regard to future events. In the context of our QP-based control framework, the reference vector field in~\eqref{eq:KCM_vecField} guides the protein towards its native conformation on the folding landscape. 
\end{itemize}

\subsection{Proof of Proposition~\ref{prop:LipCont}}

\begin{proof}
	The solution $\omega$ to the linear programming problem  
	in~\eqref{eq:unique} is the width of a feasible set associated with the QP in~\eqref{eq:ODS}. Furthermore, Lipschitz continuity of $ \mathcal{J}^\top(\pmb{\theta}) \mathcal{F}(\pmb{\theta})$  at conformation $\pmb{\theta}_0$ implies Lipschitz continuity of $\mathbf{G}(\pmb{\theta})$ in the QP. Following Theorem~1 in~[S5], which is based on Mangasarian-Fromovitz regularity conditions~[S6], uniqueness and Lipschitz continuity of the control input follow. 
\end{proof}

\subsection{Proof of Proposition~\ref{prop:trajRel}}
\begin{proof}
	Employing the control input~\eqref{eq:KCM_control} in~\eqref{eq:KCM_ode} and noting that the iteration in~\eqref{eq:KCM_alg} is the forward Euler difference equation for the closed-loop dynamics with time step $h$, the rest of the proof
	follows a standard argument from numerical analysis (see, e.g., the proof of Theorem 1.1 in~[S7]).
\end{proof}
\subsection{Details of Implementation and Computational Complexity of the Overall Algorithm}

\begin{figure}[t]
	\centering
	\includegraphics[width=0.52\textwidth]{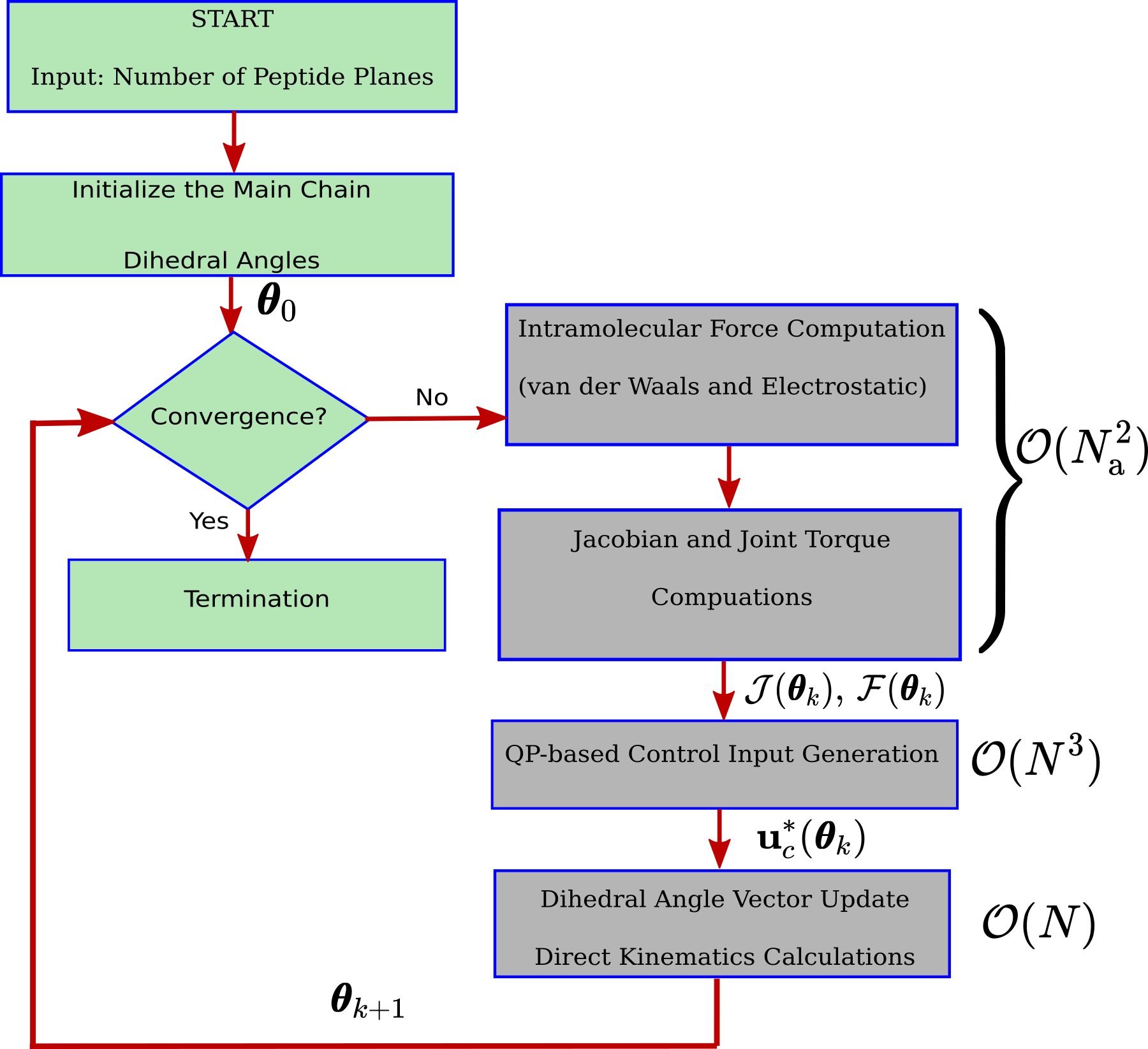}
	\caption{\small The flowchart of the computer program. Contact the corresponding author 
		(email:\texttt{amohmmad@umich.edu}) for obtaining a copy of the program.}
	\vspace{-1.5ex}
	\label{fig:flowChart}
\end{figure}

\noindent\textbf{Details of implementation.} A flowchart of the program is depicted in Figure~\ref{fig:flowChart}. The convergence 
criteria, following PROTOFOLD~I~\cite{kazerounian2005protofold}, is that the magnitude of all of the joint equivalent torques to be less 
than or equal to 1 kcal/(mol.A), which is in accordance with the experimentally determined values in folded proteins. Our QP-based control input generation block is using the MATLAB `\texttt{quadprog}' routine, where we are using the `\texttt{interior-point-convex}' as the QP numerical solver. After performing the direct kinematics calculations on the chain, we are using the `\texttt{molecule3D}' program [S9] for drawing the protein molecules. Interested readers are encouraged to contact the corresponding author (email:\texttt{amohmmad@umich.edu}) for obtaining a copy of the program.

\noindent\textbf{Computational complexity.} The big O notation for complexity analysis of numerical 
algorithms can be defined as follows. Given two real-valued functions $f(\cdot)$ and $g(\cdot)$, we say 
that $f(x)=\mathcal{O}(g(x))$ if there exists a positive real number $M$ and a real number $x_0$ such that 
$|f(x)|\leq M |g(x)|$ for all $x\geq x_0$. 

The QP given by~\eqref{eq:ODS} belongs to the class of box-constrained convex  quadratic programs that can be efficiently solved using  interior-point algorithms (see, e.g., [S8]). The time needed for solving such a QP using interior-point methods is of order $\mathcal{O}(N^3)$, where there are $N-1$ peptide planes in the protein molecule. Furthermore, if there are $N_a$ atoms in the chain, the computational complexity associated with computing the intramolecular forces at each iteration is of order $\mathcal{O}(N_a^2)$. Finally, the computational complexity associated with direct kinematics calculations is of order $\mathcal{O}(N)$. 

\noindent\textbf{Initial conformation of the protein molecule.} In the presented simulation results in 
the paper, the initial protein conformation  vector (in degrees) is given by 
\begin{align}
\nonumber
\pmb{\theta}_0 =  \big[ & 28.3, 28.6, 26.1, 28.7, 27.7, 26.0, 26.6, 27.5, 28.8, \cdots \\
\nonumber
& 28.8,  26.2,  28.8, 28.8, 27.2, 28.3, 26.2, 27.1, 28.7, \cdots \\
\nonumber
& 28.3, 28.8, 27.8, 25.8\big]^{\circ\top}.
\end{align}
\begin{figure*}[t]
	\centering
	\includegraphics[width=0.65\textwidth]{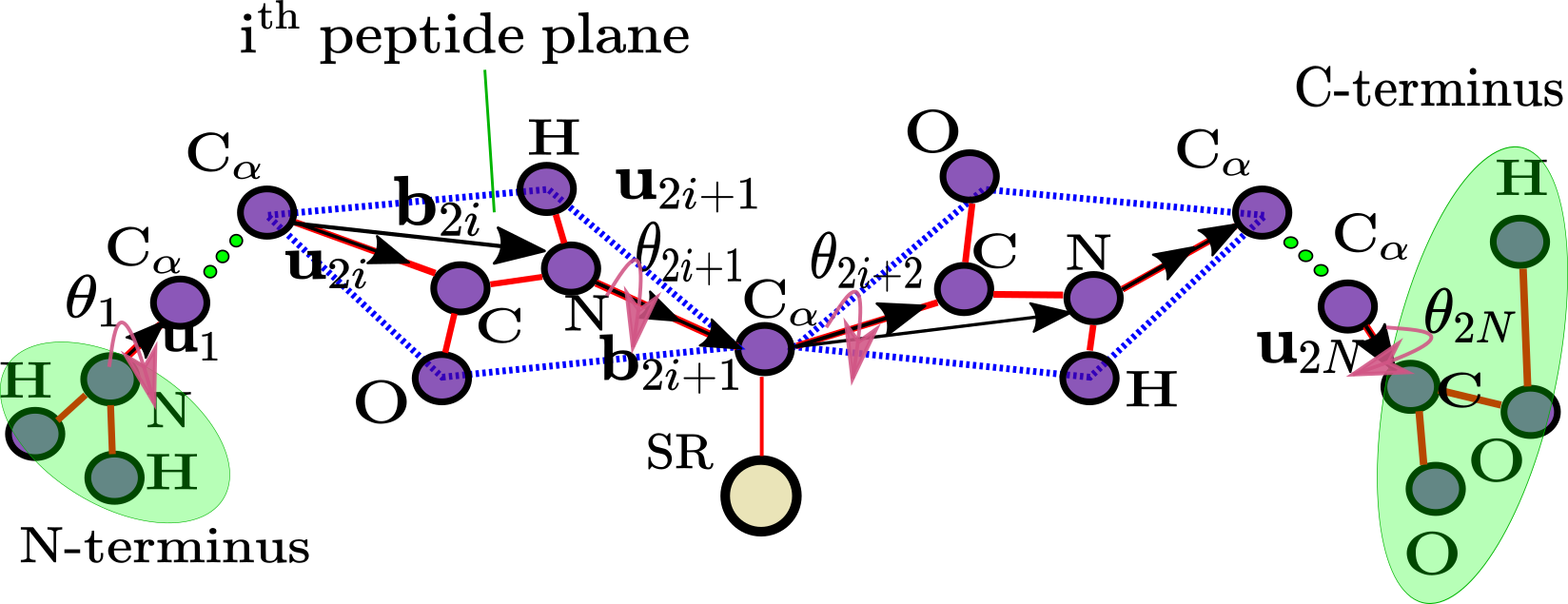}
	\caption{\small The protein molecule kinematic structure under the 
		$\text{C}_\alpha-\text{CO}-\text{NH}-\text{C}_\alpha$ coplanarity assumption.}
	\vspace{-1.5ex}
	\label{fig:protBasic1}
\end{figure*}

\section*{Supplementary References}
\small{
\noindent [S1]\hspace{1pt} S. Hikiri, T. Yoshidome, and M. Ikeguchi. ``Computational methods for configurational entropy using internal and Cartesian coordinates,'' \emph{J. Chem. Theory Comput.}, vol. 12, no. 12, pp. 5990-6000, 2016.\\

\noindent [S2]\hspace{1pt} T. Ohtsuka. ``Solutions to the Hamilton-Jacobi equation with algebraic gradients,'' \emph{IEEE Trans. Automat. Contr.}, 56(8), pp. 1874-1885, 2010.\\

\noindent [S3]\hspace{1pt} C-H. Won, S. Biswas. ``Optimal control using an algebraic method for control-affine non-linear systems,'' Int. J. Contr., vol. 80, no. 9, pp. 1491-1502, 2007.\\

\noindent [S4]\hspace{1pt} T. Cheng, F. L. Lewis and M. Abu-Khalaf. ``Fixed-final-time-constrained optimal control of nonlinear systems using neural network HJB approach,'' \emph{IEEE Trans. Neural Netw.}, vol. 18, no. 6, pp.1725-1737, 2007.\\

\noindent [S5]\hspace{1pt} B. Morris, M. J. Powell, and A. D. Ames. ``Sufficient conditions for the Lipschitz continuity of QP-based multi-objective control of humanoid robots,'' in \emph{Proc. 52nd IEEE Conf. Dec. Contr.}, Firenze, Italy, Dec. 2013, pp. 2920-2926.\\

\noindent [S6]\hspace{1pt} O. L., Mangasarian, and S. Fromovitz. ``The Fritz John necessary optimality conditions in the presence of equality and inequality constraints,'' J. Math. Anal. Appl., vol. 17, no. 1, pp. 37-47,  1967.\\

\noindent [S7]\hspace{1pt}  A. Iserles, \emph{A first course in the numerical analysis of differential
	equations}. Cambridge University Press, 2009, no. 44.\\

\noindent [S8]\hspace{1pt} Y. Wang,  and S. Boyd.  ``Fast evaluation of quadratic control-Lyapunov policy,'' IEEE Trans. Contr. Syst. Technol., vol. 19, no. 4, pp. 939-946, 2010.\\

\noindent [S9]\hspace{1pt}  A. Ludwig. \emph{molecule3D} (2016). Accessed: Oct. 2020. [Online]. Available: https://www.mathworks.com/matlabcentral/fileexchange/55231-molecule3d, MATLAB Central File Exchange. 
}

\end{document}